\documentclass[12pt]{article}
\usepackage{amsmath,amsfonts,amsthm,amssymb,amscd}
\usepackage[toc,page]{appendix}
\numberwithin{equation}{section}

\newcommand{\be}{\begin{equation}}
\newcommand{\ee}{\end{equation}}
\newcommand{\bs}{\begin{split}}
\newcommand{\es}{\end{split}}
\newcommand{\ba}{\begin{align}}
\newcommand{\ea}{\end{align}}
\newcommand{\basl}[1]{\begin{align}\begin{split}\label{#1}}
\newcommand{\bas}{\begin{align}\begin{split}}

\newtheorem{theo}{Theorem}[section]
\newtheorem{prop}[theo]{Proposition}
\newtheorem{lemm}[theo]{Lemma}

\newtheorem{defi}[theo]{Definition}

\newcommand\fpr{\hfill$\Box$\null}

\newcommand\R{\mathbb{R}}
\newcommand\C{\mathbb{C}}

\title{Markovian approximation for
Pauli Fierz operators }

\author{ L. Amour, J. Nourrigat}

\date{Laboratoire de Math\'ematiques de Reims - UMR CNRS 9008  \\ \vskip 0.15cm Universit\'e de Reims Champagne-Ardenne, France}

\begin{document}

\maketitle

\begin{abstract}
\noindent
The purpose of this article is to derive a Markovian approximation of the reduced time dynamics of observables for the Pauli-Fierz Hamiltonian with a precise control of the error terms. In that aim, we define a Lindblad operator associated to the corresponding quantum master equation. In a particular case, this allows to study the transition probability matrix.
\end{abstract}

\parindent=0pt

\


\

\parindent=0pt
\parindent = 0 cm

\parskip 5pt
\baselineskip 13pt

\section{ Introduction.}\label{s1}

     Pauli-Fierz operators are used to describe the time evolution of charged particles  with spins interacting with electric potentials and  the quantized electromagnetic field (photons) and possibly with a non quantized external magnetic field \cite{B-F-S, F73,CTD01}. In the Markovian approximation, the interaction between particles and the quantized  field is usually assumed to be small and in Pauli-Fierz Hamiltonians, the terms for these interactions are  multiplied by a parameter $g$ (coupling constant). We  then suppose in the following that $g$ is small in the Pauli-Fierz Hamiltonian considered  here and denoted by $H(g)$.

The purpose of this work is to give an approximation of a Markov type for the time evolution of  observables for small $g$. In that aim, we  shall omit some terms in the Hamiltonian that we expect would not modify our results below. The precise definition of the simplified Pauli-Fierz Hamiltonian $H(g)$ is given in Section \ref{s2}.

Let ${\cal H}_{tot}$ be the  Hilbert space of the model. Assuming that the Hamiltonian $H(g)$ is acting in  ${\cal H}_{tot}$,  we recall that the evolution $X(t, g)$ of an observable $X$ in ${\cal H}_{tot}$ is usually given for any time $t$ by:
$$ X(t, g) = e^{it H(g)} X e^{-it H(g)}. $$

In the system studied here we consider moving spinless charged particles together with a spin-$\frac{1}{2}$ fixed particle in the quantized electromagnetic field, with an external electric potential and an external constant magnetic field.
The Hilbert space ${\cal H}_{tot}$ is the completed tensor product of the three Hilbert spaces ${\cal H}_{ph}$, ${\cal H}_{el}$ and ${\cal H}_{sp}$ respectively corresponding
 to the quantized electromagnetic field,  the moving charged particles and the  spin fixed particle. Concerning the observables, we assume that they act only on the moving charged particle  and the spin fixed particle spaces and not on the photon space, that is, the observables are written as
  $I \otimes X$ with $I$ the identity in ${\cal H}_{ph}$ and $X$ an operator in
 ${\cal H}_{el} \otimes {\cal H}_{sp}$. The photon vacuum state   is denoted by  $\Psi_0\in {\cal H}_{ph}$.

 In this article we are concerned with
the operator denoted  $\sigma_0 S(t, g) X$ mapping from  ${\cal H}_{el} \otimes {\cal H}_{sp}$
into itself and defined by:
  $$ \langle \sigma_0 S(t, g) X f, g  \rangle_{{\cal H}_{el} \otimes {\cal H}_{sp}} =  \langle(I \otimes X ) e^{-it H(g) } ( \Psi_0 \otimes f),
      e^{-it H(g) }  ( \Psi_0 \otimes g)  \rangle_{{\cal H}_{tot}},$$
   for all $f$ and $g$ in ${\cal H}_{el} \otimes {\cal H}_{sp}$.
   Thus, we consider time evolutions with initial states in the vacuum of photon. Note that in the case $g=0$, one has:
   $$ \sigma_0 S(t, 0) X = e^{ it ( H_{el} + H_{sp} )} X e^{ -it ( H_{el} + H_{sp} )} $$
where $ H_{el}$ and  $H_{sp}$ respectively are  the  Schr\"odinger and the free spin  Hamiltonian
   (see Section \ref{s2.1}). That is,
   $$ \frac {d} {d t } \sigma_0 S(t, 0) X = i \sigma_0 S(t, 0)   [ (H_{el} + H_{sp}), X].  $$
This means that, for $g=0$, the observable evolution follows the  Heisenberg equation
   (for the Schr\"odinger and the free spin  Hamiltonians).

  Our goal is to prove that for $g$ small enough,  $\sigma_0 S(t, g) X $ follows a differential equation (master equation) with an explicit term (Lindblad operator) and two
   negligible terms. The first error term is estimated with ${\cal O } (g^3)$ and the second with ${\cal O } (g^2)$ but the latter can be negligible for large $t$.
   The simplified equation obtained when omitting these two error terms is the quantum master equation and constitutes the Markovian  approximation.
   Let us be more specific concerning that result. We define a mapping $L(g)$ (see Proposition \ref{defi-Lind} below)  giving for any suitable operator $X$ in ${\cal H}_{el} \otimes {\cal H}_{sp}$ an operator $L(g)X$ also in ${\cal H}_{el} \otimes {\cal H}_{sp}$ satisfying for all $t>0$:    
   $$ \frac {d} {d t } \sigma_0 S(t, g) X  =  \sigma_0 S(t, g)  L(g) X +
   R_0 (t, g)X
 + R (t, g)X.    $$
The mapping $L(g)$ is called Lindblad operator
      (see \cite{C-P,A-L,G-K-S,Haa73,Kos72,Lind,Mer19,RH12}).
Concerning the error terms in this approximation, we prove for small $g$ the following estimates for all positive times:
   $$ \Vert R_0 (t,g) X \Vert \leq C \frac { g^ 2} {1+ t} \Vert X \Vert,\quad \Vert R (t,g) X \Vert \leq C  g^3\ln(1+t)\Vert X \Vert$$
and if the ultraviolet cutoff vanishes at the origin:
   $$ \Vert R_0 (t,g) X \Vert \leq C \frac { g^ 2} {1+ t^2} \Vert X \Vert,\quad \Vert R (t,g) X \Vert \leq C  g^3\Vert X \Vert$$
   where the above norm $\Vert X \Vert$ will be specified.

  This result is precisely stated in Theorem \ref{TP1} below and is the main result of the paper. In Section \ref{s2.2}, we shall see the precise hypotheses and the definition of the chosen norms in order to get the control of the error terms.

  There exists other types of corrections of the Heisenberg equation for the Schr\"odinger operator, for example, the Breit operator (see \cite{BR,B-S}). There are other reduced dynamics, see e.g. \cite{H-S,SP}, which are different to the one that we consider.
Return to equilibrium for Pauli-Fierz are studied, e.g., in  \cite{BFSRE,D-J,DH1,DH2,DH3}.
For Markovian approximation and weak coupling limit approximation, see  \cite{Dav,SP} and also \cite{A-L,RH12}, and \cite{VH55,VH57,Dav,Dav76,RH12,SP} for the control of the error.

 Let us mention that the main result  Theorem \ref{TP1} is proved in view of possible applications in NMR (Nuclear Magnetic Resonance) for a system of several interacting electrons together with a fixed spin-$\frac{1}{2}$ particle subject to a constant external (non quantized) magnetic field. See  \cite{A-N} for the case of several fixed spins but without moving electrons and see \cite{Mer19,JP96} in the positive temperature case.

 In Section  \ref{s5}, we make explicit the operator
   $L(g)$ in a particular case. Namely, we consider the case of single electron without constant magnetic field and without the fixed spin particle. We make the hypotheses that the Schr\"odinger operator is globally elliptic in the sense of
 \cite{BH}.
In that case, there exists an Hilbertian basis  $(u_j)$ of $L^2 ( \R^3) $  with  eigenfunctions $u_j$ of the  Schr\"odinger operator $H_{el}$ (see Section \ref{s2.1} and Section \ref{s5}). Denoting by $\pi_{u_j}$ the  orthogonal projection on the  $u_j$,
   we make explicit for all  $m$ and  $j$:
$$    M _{mj } =  g^{-2}  \langle ( L(g) \pi_{u_j} ) u_m ,  u_m  \rangle. $$
  The form of this matrix suggests that it is the infinitesimal generator of a Markov semi-group.  One can address the question of the existence of this semi-group and  whether  it defines a good approximation for small $g$ of the transition probability matrix, that is, the probabilities for  an electron initially in an excited state $u_m$ without photon to be at time $t$ in the state
  $u_j$. Different facts make this realistic. First, the matrix is triangular which means that up to this approximation, the probability for the electron to be at time $t>0$ at an energy level higher than at the initial time is zero. Second, the diagonal elements of this matrix are related to the imaginary part of the resonances studied by  Bach, Fr\"{o}hlich and  Sigal \cite{B-F-S}.

\section{Statement of the results.}\label{s2}

 \subsection{ Description of the model.}\label{s2.1}

 The Hilbert space of the model is the completed tensor product of the three Hilbert spaces of the elements constituting the system:
 $$ {\cal H} _{tot} =  {\cal H} _{ph}  \otimes   {\cal H} _{el} \otimes  {\cal H} _{sp} $$
and the Hamiltonian of the model is the sum of the Hamiltonians
$H_{ph}$, $H_{el}$ and $H_{sp}$
of the three elements when they are not interacting, together with the interaction Hamiltonian $H_{int}$.

  {\it Photon Hilbert space and Hamiltonian.}

The single photon Hilbert space  ${\cal H} _{ph}^1$ is:
  $$ {\cal H} _{ph}^1 = \{ f\in  L^2 (\R^3 , \R^3 ), \quad
  \sum _{j= 1}^3 k_j f_j(k) = 0,\ k\in\R^3 \}.$$
  The photon Hilbert space   ${\cal H} _{ph}$ is the symmetric Fock space
   ${\cal F}_s ( {\cal H} _{ph}^1 )$ over the single photon Hilbert space  ${\cal H} _{ph}^1$ and we use the notations and definitions of \cite{RSII}.
   In this Fock space, we shall mainly use two types of unbounded operators. These two unbounded operators
   are initially defined in the subspace ${\cal H} _{ph}^{fin} $ of ${\cal H} _{ph}$
    of elements with a finite number of components (see \cite{RSII}). For each $V$
   in the one photon space ${\cal H} _{ph}^1$, we use the Segal field $\Phi_S (V)$ which is an
   unbounded operator in the Fock space  ${\cal H} _{ph}$ (see \cite{RSII}). For each    bounded or unbounded operator $T$
 in ${\cal H} _{ph}^1$, we use the standard operator $d\Gamma (T)$
   in the Fock space (see \cite{RSII}). In particular, if  $M_{\omega }$
   is the  multiplication in  ${\cal H} _{ph}^1$ by  $\omega (k) = |k|$,
    the operator   $ H_{ph} = d\Gamma ( M_{\omega }) $  is the usual free photon energy Hamiltonian.

{\it Electron Hilbert space and Hamiltonian.}

The electron Hilbert space is  $  {\cal H} _{el} = L^2 (\R^{3Q})$ where $Q$ is the number of electrons.
We assume that there is a constant external  magnetic field  $B_{ext } = (0, 0, B)$ where $B\in\R$. The corresponding vector potential is denoted by:
$$  A^{ext}(x) = \left( - B \frac {x_2} {2} ,  B \frac {x_1} {2} , 0\right).$$
Letting $x = ( x^{(1)} , \dots, x^{(Q)}) $ be the variable in $\R^{3Q}$, we set for $\alpha=1,\dots,Q$ and $j=1,2,3$:
$$ D_j^{(\alpha )} = \frac {1} {i} \ \frac {\partial } {\partial x _j^{(\alpha )}  },\quad
 \nabla _j ^{(\alpha )}  =   D_j^{(\alpha )}  - A_j^{ext}(x^{(\alpha )}).
$$
The Hamiltonian of the system of electrons is the $Q$-body
 Schr\"odinger
 operator with constant magnetic field  $(H_{el }, D(H_{el }) ) $:
  \be\label{Hel}  H_{el} = \sum _{\alpha =1}^Q   \sum _{j=1}^3
  (  \nabla _j ^{(\alpha )}  )^2   + V(x) \ee
 where $V$ is the electric potential supposed to be  a  real valued function on $\R^{3Q}$ with polynomial growth at infinity and
 identified with   the multiplication
  operator by this function. This Hamiltonian operator is initially defined on
  ${\cal S} (\R^{3Q })$. One can also omit the polynomial growth on the electric potential but replace in the following the Schwartz space by the  set of $C^\infty$ functions with compact support

{\it Spin Hilbert space and Hamiltonian.}

The spin Hilbert space of the fixed particle  is $  {\cal H} _{sp} = \C^2$.
We denote by  $\sigma _j$ the standard Pauli matrices ($1 \leq j \leq 3$). In the case of a constant magnetic field
  $(0, 0, B)$, the energy of the spin-$\frac{1}{2}$ particle corresponds to the Hamiltonian:
  \be\label{Hsp}  H_{sp} = B  \sigma _3. \ee

 {\it Interaction Hamiltonian.}

The interaction between  electrons and the spin fixed particle with the quantized electromagnetic field is expressed with two types of operators. One type corresponds to  the three components of the quantized electromagnetic vector potential and is used to define the interaction between the electron motions and the field. The other type, corresponding to the three components of the quantized magnetic field at the position $x_0$ of the spin-$\frac{1}{2}$ fixed particle, is used to define the interaction between the field and the spin  particle.
These operators are  the image  by the Segal field  $\Phi_S$ of the elements  $ A_{jx} $ and $B_{jx}$ of ${\cal H} _{1}$  ($x\in \R^3$,  $1\leq j\leq 3$)  that we recall here the definitions:
  \be\label{A-j}  A_{jx}(k)  = \frac { \varphi (|k|) } {|k|^{1/2}} \ e^{ - i k\cdot x } \
 \left ( e_j - (e_j \cdot k) \frac { k } {|k|^{2}} \right ),\quad k\in \R^3\backslash\{0\},  \ee
\be\label{B-j} B_{j,x }(k) = {i\varphi(|k|)|k|^{1\over 2} \over (2\pi)^{3\over 2}}
  \ e^{ - i k\cdot x } \  {k\times e_j\over |k|},\quad k\in\R^3\backslash\{0\}.\ee
In the above equalities,  $(e_1 , e_2, e_3)$ stands for the canonical basis of  $\R^3$.
The function  $\varphi$ is a smooth ultraviolet cutoff always supposed in the Schwartz space ${\cal S}(\R)$.
Note that each $ A_{jx}$ and $ B_{jx}$ ($j$ and $x$ fixed) maps $\R^3$ into $\R^3$ and satisfy the equalities:
$$ k \cdot A_{jx}(k) = k \cdot B_{jx}(k) = 0, \quad k\in\R^3\backslash\{0\}.$$
This means that the $A_{j,x }$ and $B_{j,x }$ are elements of the
single photon Hilbert espace ${\cal H} _{ph}^1$ according to the decreasing property of  $\varphi$.
The operators corresponding to the three components of the vector potential and of the magnetic field at point $x$ respectively are  $\Phi_S (  A_{jx})$
and  $\Phi_S (  B_{jx})$.

In order to define and use for upcoming computations the operator $ H_{int} $, we temporarily identify
         ${\cal H} _{tot} $ and  $L^2 (\R^{3Q}  , {\cal H} _{ph}  \otimes {\cal H} _{sp} )$ and set, for all $f\in  {\cal S}(\R^{3Q} , {\cal H} _{ph}  \otimes {\cal H} _{sp} )$
and any  $x\in \R^{3Q}$:
   $$ ( H_{int} f ) (x)  = \sum _{\alpha = 1}^Q \sum _{j=1} ^3
    \Big ( \Phi_S ( A _{j, x^{[\alpha]} } ) \otimes I \Big )
      \nabla _j ^{(\alpha )}  f(x) + \sum _{j=1} ^3  \Big (  \Phi_S ( B _{j, x^{[0]} } ) \otimes \sigma _j \Big )
     f(x).$$
     The tensor product in the above equality refers to
     $ {\cal H} _{ph}  \otimes {\cal H} _{sp} $ and $I$ denotes the identity in $ {\cal H} _{sp} $. It appears to be more convenient in the sequel to use  integral expressions with creation operators $a^{\star } (k)$ and annihilation operators  $a(k)$ ($k\in  \R^3$) as in \cite{B-F-S}.
For any  $V\in {\cal H} _{ph}^1 $:
       $$ \Phi_S (V) = \frac {1 } {\sqrt 2 }  \int _{\R^3 }
       \Big ( a^{\star } (k)  V(k) + a(k) \overline {V(k) } \Big ) dk. $$
Then for each $k\in \R^3$, we define an operator $E(k)$
      from $ {\cal H} _{el}  \otimes {\cal H} _{sp} $
       now identified with   $L^2 (\R^{3Q}  , {\cal H} _{sp} )$ and taking values in
        $ ({\cal H} _{el}  \otimes {\cal H} _{sp} ) ^3 $ by:
     \be\label{def-Ek} (E(k) f) (x)  = \sum _{\alpha = 1}^Q \sum _{j=1} ^3  A _{j, x^{[\alpha]} } (k)
        \nabla _j ^{(\alpha )}  f(x) +  \sum _{j=1} ^3  B _{j, x^{[0]}}  (k) \sigma _j f(x)  \ee
for all $f\in  {\cal S}(\R^{3Q} , {\cal H} _{sp} )$.

  The operator $E^{\star} (k) $ denotes the formal adjoint of $E(k)$.  Let us emphasize that  each $E(k)$ takes values  in  $ ({\cal H} _{el}  \otimes {\cal H} _{sp} ) ^3 $ and verifies:
   $$ k\cdot E(k) = 0,\quad  k\in \R^3.$$
    Therefore, for each $f\in  {\cal S}(\R^{3Q} , {\cal H} _{sp} )$
    and for each $x\in\R^{3Q}$, the function $k \rightarrow E_x (k) = (E(k)f) (x)$ belongs to $ {\cal H} _{ph}^1 \otimes {\cal H} _{sp}$  where  $ {\cal H} _{ph}^1 $ is
    the single photon Hilbert space. Thus, extending the Segal field $\Phi_S$ to
    the Hilbert space  $ {\cal H} _{ph}^1 \otimes {\cal H} _{sp}$ as in \cite{DG},
    we have:
     $$ ( H_{int } f) (x) = \Phi_S (E_x).$$
     This operator is initially defined in ${\cal H }_{tot } ^ {reg } = {\cal H }_{ph } ^ {fin}
     \otimes {\cal S} (\R^{3Q}) \otimes {\cal H }_{sp }$.  This definition may   be written
     in a somewhat formal way but useful in the following sections as:    
\be\label{Hint-Ek}  H_{int} =  \frac {1} {\sqrt 2} \int _{\R^3}  \Big ( ( a(k)  \otimes E^{\star} (k)  + a^{\star} (k)   \otimes E(k)\Big )  dk. \ee
   The above tensor product corresponds to the tensor product of  ${\cal H }_{ph }$   with
   ${\cal H} _{el}  \otimes {\cal H} _{sp}$.

  {\it Full Hamiltonian.}

  The following operators are initially defined in
  ${\cal H }_{tot } ^ {reg } $:
   \be\label{def-H-0}  H(0)= H_{ph} + H_{el} + H_{sp},\ee
  $$ H(g) =  H_{ph} + H_{el} + H_{sp} + g H_{int} = H(0) +  g H_{int} $$

{\it Domain and unitary group.}

The assumption on the Hamiltonian   $H_{el}$
is hypothesis  $(H) $ below. We use the following notations. For any  $( \alpha , J)$ with  $\alpha =
(\alpha _1 , \dots, \alpha _m)$ and
$J = ( j_1 , \dots, j_m)$ where the  $\alpha _{\mu}=1,\dots, Q$
and the  $j_{\mu}=1,2,3$,
we set:
$$ D_{\alpha  , J} = \nabla _{j_1}  ^{[\alpha_1  ]} \cdots
\nabla _{j_m}  ^{[\alpha_m  ]} $$
and  set $|(\alpha , J)|  = m$. We make the following hypothesis $(H)$.

{\it
{\bf Hypothesis {\it (H)}.} The operator $H_{el}$ defined in (\ref{Hel}) is essentially self-adjoint and we denote by  $H_{el}$ its self-adjoint extension. We assume that there exists  $C>0$  satisfying $H_{el} + C >0$ and we denote by $W_m^{el}$ the domain of  $( C+ H_{el})^{m/2}$.
For all integers $m$,  $W_{m}^{el}$  is the set of all  $u\in L^2(\R^{3Q})$ satisfying $ D_{\alpha  , J} u\in L^2(\R^{3Q})$ if $|(\alpha , J)|  \leq  m$ with the norm:

 \be\label{def-Dom}  \Vert u \Vert _{W_{m}^{el}} ^2 = \Vert u \Vert  ^2  +
   \sum _{| (\alpha , J )| \leq m }
  \Vert D_{\alpha  , J}  u \Vert ^2  \ee
where $\Vert \, \cdot\,\Vert$ is the $L^2(\R^{3Q})$ norm.}

 This hypothesis is fulfilled  for some potentials $V$. In particular, we consider in Section \ref{s5} the case of globally elliptic Schr\"odinger operator (\cite{BH}) where this hypothesis is satisfied. Our main result could be applied with other potentials.

Since we consider $H_{el}+H_{sp}$ with $H_{sp}$ bounded we shall still denote by $W_{m}^{el}$ the spaces associated to $H_{el}+H_{sp}$ instead of $H_{el}$ in the aim to avoid additional notations such as $W_{m}^{el-sp}$. For example, $W_{0}^{el}$ now refers to ${\cal H}_{el}\otimes {\cal H}_{sp}$.

 \begin{theo}\label{domain}   Under hypothesis  $(H)$, the operator $H(0)$ initially defined on  ${\cal H}_{tot}^{reg}$ is essentially self-adjoint. Let $C>0$ be such that $H(0) + C >0$ and let
 $ W_m ^{tot}$ be the domain of $ (H(0) + C)^{m/2}$. Then, for $g$ small enough,
 the operator $H(g)$ with domain  $ W_2 ^{tot}$ is self-adjoint and the domain of
  $ (H(g) + C)^{m/2}$ is $ W_m ^{tot}$.

  \end{theo}
  Theorem \ref{domain} is proved in Section \ref{s3}.
 By this theorem, the operators $e^{it H(0)}$ and $e^{it H(g)}$ make sense for
   $g$ small enough.  See also \cite{H02} for self-adjointness of Pauli-Fierz without the condition $g$ small.

 \subsection{ Observable time evolutions.}\label{s2.2}

{\it Full, free and reduced time evolutions of    electron-spin observable.}

For every operator  $X$ in ${\cal H}_{el}\otimes {\cal H}_{sp}$, the  time full evolution of $X$ is defined by:
$$ S(t,g) X = e^{it H(g) }  ( I_{ph}   \otimes X)  e^{-it H(g) }.$$
For any operator $A$ in ${\cal H} _{tot}$, we denote by
$\sigma _0 A $ the operator defined in ${\cal H} _{el} \otimes {\cal H} _{sp} $ by:
\be\label{sigma-0}   \langle \sigma _0 A f , g  \rangle _{{\cal H} _{el} \otimes {\cal H} _{sp}  }
=  \langle  A ( \Psi_0 \otimes  f ), ( \Psi_0 \otimes  g )  \rangle_{{\cal H} _{tot}  },  \ee
for all  $f$ and $g$ in ${\cal H} _{el} \otimes {\cal H} _{sp} $.

In particular, we have for $S(t, g)X$, $t>0$:

\begin{defi} We call reduced time evolution the mapping $X \rightarrow \sigma _0 S(t,g)X$ from ${\cal H}_{el} \otimes {\cal H}_{sp}$ into ${\cal H}_{el} \otimes {\cal H}_{sp}$ defined for every  $t>0$ and any operator $X$ in ${\cal H}_{el} \otimes {\cal H}_{sp}$ by: 
$$  \langle \sigma _0 S(t, g)X f , g  \rangle _{{\cal H} _{el} \otimes {\cal H} _{sp}  } =  \langle (I_{ph} \otimes X)
  e^{-it H(g) } ( \Psi_0 \otimes f ) , e^{-it H(g) } ( \Psi_0 \otimes g )  \rangle_{{\cal H}_{tot}},$$
for all $f$ and $g$ in ${\cal H}_{el}  \otimes {\cal H}_{sp}$.

 \end{defi}

  In rest of the paper, we often omit the Hilbert space as a subscript in $\Vert \cdot \Vert$ and $ \langle\, ,\, \rangle$  when it is ${\cal H}_{tot}$, ${\cal H}_{ph}$, ${\cal H}_{el}$ or ${\cal H}_{el}\otimes {\cal H}_{sp}$. Similarly we omit the Hilbert subscript in the notation $I$ for the identity since there is no ambiguity. Also ${\cal L}({\cal H}_1,{\cal H}_2)$ denotes the set of bounded linear maps from ${\cal H}_1$ to ${\cal H}_2$ for any Hilbert spaces ${\cal H}_1$ and ${\cal H}_2$.

The free evolution of an  operator $A$ in  ${\cal H}_{tot}$ is defined by:
 \be\label{gamma-t-free}   A ^{free} (t) = e^{it  H(0) }
 A  e^{- it H(0)   } \ee
and  for an operator $A$ in ${\cal H}_{el} \otimes {\cal H}_{sp}$, we set (using the same notation since no confusion will appear):
 $$ A ^{free}  ( t) =  e^{  it ( H_{el} + H_{sp} ) } A e^{  - it ( H_{el} + H_{sp} ) }. $$
In particular, with  $E(k)$ the function used to give an expression of  $H_{int}$ in
(\ref{Hint-Ek}), we set:
 \be\label{E-k-t} E ^{free}  (k, t) =  e^{  it ( H_{el} + H_{sp} ) } E(k) e^{  - it ( H_{el} + H_{sp} ) }   \ee
and $ E^{free, \star} (k, t)$ is similarly defined.

{\it Definition of the Lindblad operator.}

    The  Lindblad operator is  defined by the following Proposition proved in Section \ref{s4}.
    
Let us recall that the operator $E(k)$ defined in (\ref{def-Ek}) is an operator
from $ {\cal H} _{el}  \otimes {\cal H} _{sp} $ to $ ({\cal H} _{el}  \otimes {\cal H} _{sp} ) ^3 $.
Therefore, we can write $E(k) = ( E_1 (k), E_2 (k) , E_3 (k))$. In this subsection, we agree in order to avoid running indices in the sums to write $E^{\star } (k) A E(k)$
instead of   $\sum _j E_j^{\star } (k) A E_j(k)$  for any operator $A$ in all the integrals.

    \begin{prop}\label{defi-Lind}  Let  $X$ be a bounded operator in  ${\cal H}_{el} \otimes {\cal H}_{sp}$ and bounded from
 $W_1^{el} $ into itself. Then, for every  $g$ small enough and  all $t>0$,
 the following operator is well defined from $W_2^{el}$
into $W_0^{el} = {\cal H}_{el} \otimes {\cal H}_{sp} $:
 \be\label{L-1-t}  L(t, g ) X = i [ H_{el} + H_{sp}  , X ] + $$
 $$   \frac {g^2 }   {2}  \int _{\R^3 \times (0, t) }  \left ( e^{ - i s |k |}
  [ E^ {\star } (k), X]\  E^{free} (k, - s)
-  e^{  i s |k |} E^ {free} ( k, -s) ^{ \star }  \  [ E(k), X]  \right ) dk  ds.   \ee
Moreover, for small $g$, the limit as $t\to +\infty$ of $L(t,g)X$ exists in  ${\cal L} ( W_2^{el} , W_0^{el})$
     and this limit is denoted  by $L(g) X$. We also have for some $C>0$:
    $$ \Vert L(t, g)X - L(g) X \Vert _{ {\cal L} ( W_2^{el} , W_0^{el}) }
    \leq \frac { C g^2  } { 1+t } \Vert X \Vert_{ {\cal L} ( W_1^{el} , W_1^{el}   ) },  $$
   for all $t>0$.
   If the ultraviolet cutoff function $\varphi$ in  (\ref{A-j})(\ref{B-j})
 vanishes at the origin then this inequality can be replaced by: 
    $$ \Vert L(t, g)X - L(g) X \Vert _{ {\cal L} ( W_2^{el} , W_0^{el}) }
    \leq \frac { C g^2 } { 1+t^2 } \Vert X \Vert_{ {\cal L} ( W_1^{el} , W_1^{el}   ) }. $$

       \end{prop}

The operator $L(g):$ $X\to L(g)X$ is called Lindblad operator.

{\it Main result.}

\begin{theo}\label{TP1} Let $X$ be an operator in $W_0^{el} = {\cal H} _{el} \otimes {\cal H} _{sp}$
bounded in $ W_j^{el}$ for all $0\leq j\leq 4$.  We have:
 \be\label{equa-diff} \frac {d} {dt}\sigma _0 S(t,g) X = \sigma _0 S(t,g) L(g) X + R_0 (t,g) X
 + R (t,g) X   \ee
where  $R_0(t,g) X$ is an operator from  $W_2 ^{el} $ into $W_0 ^{el} $ and $R(t) X$
is an operator from $W_4 ^{el} $ into $W_0 ^{el} $ satisfying:
 $$ \Vert R _0(t,g) X \Vert_{ {\cal L} ( W_2^{el}  , W_0 ^{el} )}   \leq  C \frac { g^ 2} {1+ t}
 \Vert X \Vert _{ {\cal L} ( W_1^{el} , W_1^{el}  )}     $$
 $$ \Vert R (t ,g) X \Vert_{ {\cal L} ( W_4^{el} , W_0 ^{el} )}  \leq C g^3 \ln (1+t)
  \sup _{m\leq 4}  \Vert X \Vert _{ {\cal L} ( W_m ^{el} , W_m ^{el}  )}.  $$
If the ultraviolet cutoff $\varphi$ in (\ref{A-j})(\ref{B-j})
 vanishes at the origin then these inequalities can be improved by: 
 $$ \Vert R _0(t ,g) X \Vert_{ {\cal L} ( W_2^{el}  , W_0 ^{el} )}   \leq  C \frac { g^ 2} {1+ t^2}
 \Vert X \Vert _{ {\cal L} ( W_1^{el} , W_1^{el}  )}     $$
 $$ \Vert R (t ,g) X \Vert_{ {\cal L} ( W_4^{el} , W_0^{el} )}  \leq C g^3
  \sup _{m\leq 4}  \Vert X \Vert _{ {\cal L} ( W_m ^{el} , W_m ^{el}  )}.  $$

\end{theo}
Theorem \ref{TP1} is proved in Section \ref{s4}.

\section{  Sobolev spaces. Proof of Theorem \ref{domain}.}\label{s3}

Theorem \ref{domain}  follows from Proposition \ref{self-adj-H0}
 and Proposition \ref{self-ad-Hg}.

\subsection{ Electronic Hamiltonian.}\label{s3.1}

We are first concerned with studying the action of the operator $E(k)$ defined in (\ref{def-Ek}) in the Sobolev spaces
 $W_m^{el}$.   Spherical coordinates  for $k$, namely $k= \rho \omega$
($\rho >0$ and $\omega \in S^2$ the unit sphere centered at the origin) are used in that purpose.

\begin{prop}\label{act-deriv-E-k}   Under hypothesis $(H)$ the operator  $E(k)$ defined in (\ref{def-Ek}) for fixed $k\in\R^3$   maps
  $W_m^{el}$ into $(W_{m-1}^{el}) ^3$. Moreover, for any $\alpha $, all $m$ and $p$, there is $C_{\alpha m p }>0$ such that:
  \be\label{deriv-E-k} \Vert \partial _{\rho}^{\alpha }  \left(    \rho ^{1/2}   E   (\rho \omega ) \right )
      \Vert  _{ {\cal L} ( W_m^{el} , (W_{m-1}  ^{el})^3 ) } \leq C_{\alpha m p } ( 1+ \rho)^ {-p}. \ee
In particular, there exists $C_{ m p }$ satisfying:
 $$  \Vert    E   (k )
      \Vert  _{ {\cal L} ( W_m^{el} , (W_{m-1}  ^{el})^3 ) }  \leq C_{ m p } ( 1+ |k| )^ {-p}. $$
The analogous estimates holds for $E^{\star} (k)$.

  \end{prop}

 {\it Proof.} We use hypothesis (\ref{def-Dom}) noticing according to the expressions (\ref{A-j}) and (\ref{B-j}) for the functions $A_{jx}$ and $B_{jx}$ that the commutator:
 $$        [  \nabla _{j}  ^{[\alpha  ]}   , E(k) ] = i k_j E(k). $$
Since the smooth ultraviolet cutoff $\varphi$ in
 (\ref{A-j})(\ref{B-j}) is supposed to belong to ${\cal S} ( \R)$, we see that for each integer $p$:
$$ \Vert \partial _{\rho}^{\alpha }  \left (    \rho ^{1/2}   E   (\rho \omega ) \right )
      \Vert  _{ {\cal L} ( W_1^{el} , (W_{0}  ^{el})^3 ) } \leq C_{\alpha m p } ( 1+ \rho)^ {-p}.  $$
      The proposition follows. \fpr

\subsection{ Free Hamiltonian.}\label{s3.2}

We first have the following property (\cite{PUT}).

 \begin{prop}\label{self-adj-H0}   The operator $H(0)$ defined in ${\cal H} _{tot} ^{reg}$  by (\ref{def-H-0})
  is essentially self-adjoint. We also denote by  $H(0)$ its self-adjoint extension. If $C>0$ satisfies
 $H(0) + C >0$ then $W_m^{tot}$ stands for the domain of $( C+ H(0))^{m/2}$. For all integers $m$,
we have:
  \be\label{caract-W-tot}  \Vert f \Vert _{ W_{2m}^{tot}} ^2 = \sum _{p+q \leq m}
 \Vert H_{ph} ^p H_{el} ^q f \Vert ^2. \ee

  \end{prop}

 We then deduce the Proposition below.

     \begin{prop}\label{norme-tenso} We have the two following properties.

     (i)  If the operator  $A$ maps $W_m^{el}$ into $W_p^{el}$
   then $I \otimes A$ maps $W_m^{tot}$ into $W_p^{tot}$.

     (ii)  If the operator $B$ maps $W_m^{tot}$ into  $W_p^{tot}$
  then  $\sigma _0 (B)$ maps $W_m^{el}$ into  $W_p^{el}$.

      \end{prop}

 {\it Proof.}
    Let us prove the  point $(ii)$.  We see that $ ( C+ H_{el })^p
      \sigma _0 (B) =
      \sigma _0 ( ( I \otimes  ( C+ H_{el })^p) B ) $. Therefore,
      for each $f$ in $W_m^{el}$:
      $$ \Vert \sigma _0 (B)f \Vert _{W_p^{el}} \leq
      \Vert ( I \otimes ( C+ H_{el })^p ) B(\Psi_0 \otimes f)  \Vert
   \leq \Vert  B (\Psi_0 \otimes f )   \Vert _{W_p^{tot}} $$
   $$  \leq K  \Vert \Psi_0 \otimes f  \Vert _{W_m^{tot} } =
   K   \Vert \Psi_0 \otimes H_{el }^m f  \Vert =
    \Vert  f  \Vert  _{W_m^{el} },   $$
      for some $K>0$.
  We have used above that $H_{ph} \Psi_0 = 0$.

   \fpr

We shall also use the fact that
 $e^{it H(0) }$ is bounded from $W_m^{tot}$  into itself for all $m$.

\subsection{ Full Hamiltonian.}\label{s3.3}

We now study the action of the operator  $H_{int}$ in the  Sobolev spaces defined in Section \ref{s3.2}.

 \begin{prop}\label{H-int-Sobol}  The operator  $H_{int}$ is bounded from $W_m^{tot}$ to
 $W_{m-2}^{tot}$.

\end{prop}

 For the proof, we need the following Lemma which is Lemma I.6 of \cite{B-F-S}.

\begin{lemm}\label{lemma-BFS}  Let   $k \rightarrow F(k)$  be a  function on  $\R^3$ taking values in
   ${\cal L} ( {\cal H} _ {el} \otimes {\cal H} _ {sp}, ( {\cal H} _ {el} \otimes {\cal H} _ {sp})^3 )$ and satisfying  $k\cdot F(k) = 0$ for all  $k\in \R^3$. We define two operators $T_F$ and $U_F$ from
   $W _1^{tot}$ to  $W _0^{tot}$ by:
   $$ T_F f = \int _{\R^3 }  ( a(k)\otimes F^{\star }(k) ) f dk, $$
   $$ U_F f = \int _{\R^3 }  ( a^{\star } (k)\otimes F (k) ) f dk $$
    (with the same notation convention as for $E(k))$. Then:
   $$ \Vert T_Ff \Vert^2  \leq \Vert  ( H  _{ph} ^{ 1/2} \otimes I)  f \Vert ^2
   \int _{\R^3 } \Vert F^{\star } (k) \Vert^2
   \frac {dk} {|k|},$$
   $$ \Vert U_Ff \Vert^2  \leq \Vert  ( H  _{ph} ^{ 1/2} \otimes I)  f \Vert ^2
   \int _{\R^3 } \Vert F(k) \Vert ^2
   \frac {dk} {|k|} + \Vert    f \Vert ^2
   \int _{\R^3 } \Vert F(k) \Vert^2 dk.  $$
   The norms of $F(k)$ and $F^{\star} (k) $ above are the norm of
   ${\cal L} ( {\cal H} _ {el} \otimes {\cal H} _ {sp}, ( {\cal H} _ {el} \otimes {\cal H} _ {sp})^3 )$.

  \end{lemm}

 {\it Proof of  Lemma \ref{lemma-BFS}.} For the convenience of the reader we recall the proof in \cite{B-F-S}. We have,  for all $f$ and $g$:
 $$ |  \langle T_F f  , g  \rangle | ^2 \leq \int _{\R^3 } |k| \Vert ( a(k) \otimes I) f \Vert ^2 dk \
  \int _{\R^3 }  \Vert ( I \otimes F^{\star } (k)) g \Vert ^2 \frac {dk} {|k|}. $$
  We know that:
  $$ \int _{\R^3 } |k| \Vert ( a(k) \otimes I) f \Vert ^2 dk  \leq
  \Vert  ( H  _{ph} ^{ 1/2} \otimes I)  f \Vert ^2. $$
  We also have:
  \begin{align*}
 \Vert U_Ff \Vert ^2 &= \int _{\R^6}   \langle  ( a(p) a^{\star }(k) \otimes F(k) ) f,
   (I \otimes F(p) f \rangle dk dp \\
&= \int _{\R^3} \Vert (I \otimes F(k)) f \Vert ^2 dk +
   \int _{\R^6}   \langle ( a^{\star }(k) a(p)  \otimes F(k) ) f,
   (I \otimes F(p)) f  \rangle dk dp \\
& \leq \int _{\R^3} \Vert (I \otimes F(k)) f \Vert ^2 dk +
   \int _{\R^6} \frac {|p|}  {|k|}   \Vert ( a(p)  \otimes F(k) ) f \Vert ^2 dk dp \\
& \leq \int _{\R^3} \Vert (I \otimes F(k)) f \Vert ^2 dk +
    \int _{\R^3} \frac {1}  {|k|}  \Vert ( H_{ph} ^{1/2}   \otimes F(k) ) f \Vert ^2 dk.
       \end{align*}

  The Lemma follows.\fpr

 {\it Proof of  Proposition \ref{H-int-Sobol}.} We use the identity (\ref{caract-W-tot}) for $W_m^{tot}$ together with the integral expression (\ref{def-Ek})  for $H_{int}$. We know that:
  $$  H_{ph}  a(k) =  a(k) ( H_{ph} - |k| ).   $$
  Therefore, for all $f$ in $W_{2m}^{tot}$, for all integers  $p$
  and $q$ such that $p+q \leq m-1$,
 $$  ( H_{ph} ^p \otimes H_{el} ^q ) ( a(k) \otimes E^{\star } (k))  f  =
  \sum _{j= 0}^p  (-1)^j C_p^j  |k| ^{p-j} (a(k)  H_{ph} ^ j \otimes  H_{el} ^q E(k)   ) f.$$
  We apply Lemma \ref{lemma-BFS} with:
  $$ F_j (k) = |k| ^{p-j}  H_{el} ^q E(k) ( C + H_{el} ) ^{ -q-1},\quad
  g_j  = (  H_{ph} ^ j \otimes   ( C +  H_{el} ) ^{q+1} ) f. $$
  Thus:
  $$  ( H_{ph} ^p \otimes H_{el} ^q ) ( a(k) \otimes E^{\star } (k))  f  =
  \sum _{j= 0}^p  (-1)^j C_p^j T_{F_j} g_j. $$
  By Proposition \ref{deriv-E-k}:
   $$  \int _{\R^3} \Vert F_j  (k) \Vert  ^2
   \frac {dk} {|k|} < \infty.  $$
   The norm for $\Vert F_j  (k) \Vert$ in the above integral is the norm of ${\cal L} ( {\cal H} _ {el} \otimes {\cal H} _ {sp}, ( {\cal H} _ {el} \otimes {\cal H} _ {sp})^3 )$. By the  Lemma:
     $$ \Vert ( H_{ph} ^p \otimes H_{el} ^q ) ( a(k) \otimes E^{\star } (k))  f  \Vert
     \leq C \sum _{j= 0}^p   \Vert    ( H  _{ph} ^{ 1/2} \otimes I)  g_j \Vert $$
  $$ \leq C \sum _{j= 0}^p   \Vert    ( H  _{ph} ^{ j + 1/2} \otimes
  ( C +  H_{el} ) ^{q+1}   )   f  \Vert   \leq C \Vert f \Vert _{W_{2m }^{tot }}. $$
Proposition \ref{H-int-Sobol} then follows for  $m$ even for other $m$ by
  interpolation. \fpr

 \begin{prop}\label{self-ad-Hg}   For $g$ small enough, $H(g)$ is essentially self-adjoint on its initial domain. The domain of its self-adjoint extension is  $W_2^{tot }$. The domain of  $ H(g)^m $ is $W_{2m }^{tot }$ for any integer $m$. The operator $e^{it H(g)} $ is bounded in $W_{m }^{tot }$ uniformly in $t$.

   \end{prop}

   {\it Proof.}  By Proposition \ref{H-int-Sobol}, we have
   for $g$ small enough:
 $$ \Vert H(g) ^m f - H(0) ^m f \Vert  \leq C g  \Vert  f \Vert
 _{W_{2m }^{tot }}.  $$
    Since  $W_{2m }^{tot }$ is the domain of the self-adjoint extension of $H(0)^m$,
     the Proposition follows from Kato Rellich Theorem. The last point of the Proposition is deduced from the first point for even $m$ since  $e^{it H(g)} $
 maps $ D( H(g)^{m/2}) $ into itself and  by interpolation for the other
     $m$. \fpr

\section{ Proof of the main results: Proposition \ref{defi-Lind} and Theorem \ref{TP1}.}  \label{s4}

\subsection{ Formal identities.}\label{s4.1}

We write
    $A \sim B$
  if  $\sigma _0 (A-B)= 0$ for two given operators $A$ and $B$ in ${\cal  H } _{tot}$  where $\sigma _0$ is the operator defined in  (\ref{sigma-0}).

We begin this subsection with formal equalities and the norm estimates making sense for operators will be studied in the next subsections.

For each $t>0$, set:
\be\label{E-t} E(t) = \{ (s_1 , s_2) \in \R_+^2 , \   s_1 + s_2 < t \}. \ee

\begin{theo}\label{equation +reste} For each operator $X$ in ${\cal H}_{el } \otimes {\cal H}_{sp } $,
we have:
$$ \frac {d} {dt }  e^{ it H(g) }  (I \otimes X) e^{ - it H(g) } \sim
 e^{ it H(g) } ( I \otimes L(t, g) X) e^{ - it H(g) } + R_1(t, g)X + R_2(t, g)X $$
 where  $L(t, g) X$ is defined in   (\ref{L-1-t}) and:
 \be\label{R1-t} R_1(t, g)X = i \frac {g^3   }   { 2 } \int _{\R^3 \times E(t) } e^{ i (t - s_1) |k| }
 e^{ it H(g) } \ \Big ( I \otimes [ E^{\star } (k), X ] \Big )
e^{ -i s_2 H(g)  }  \ee
$$  [ H_{int} , I \otimes E ( k, s_1 + s_2 - t ) ]
e^{ i (s_2 - t)  H(g)  } dk ds_1 ds_2, $$
\be\label{R2-t}  R_2 (t,g) X = - i \frac {g^3   }   { 2 } \int _{\R^3 \times E(t) }
   e^{ i ( s_1 - t) |k| }  e^{ i (t - s_2) H(g) }
    [ H_{int} , I \otimes E^ {free}  ( k, s_1 + s_2 - t )^{\star}  ] \ee
    $$
   e^{ i  s_2 H(g) } ( I \otimes [ E (k), X ]  )  e^{ - i  t H(g) }
   dk ds_1 ds_2.  $$

\end{theo}

Concerning the dependance on the parameter $g$ in the notations, we make it explicit in important terms such as $H(g), S(t,g), L(t,g), L(g), R_0(t,g), R(t,g),R_1(,g), R_2(t,g)$ and not  for the terms used in the computations such as $f(t), I_1(t), I_2(t), L_1(t),L_2(t),\Phi(s,X)$.

The first three steps of the proof correspond to the three Propositions below.

\begin{prop}\label{step-1} For all operators $X$ in  ${\cal H}_{el } \otimes {\cal H}_{sp } $,
we have:
$$ \frac {d}  {d t } e^{ it H(g) }  ( I \otimes X)
 e^{ - it H(g) } = i e^{ it H(g) } \Big ( I \otimes  [ (H_{el} +  H_{sp}), X ]  +
 g [  H_{int}, I \otimes X ] \Big )    e^{ - it H(g) }  $$
 $$ = i e^{ it H(g) }  ( I \otimes  [ (H_{el} +  H_{sp}), X ] ) e^{ it H(g) } $$
 $$  + \frac { ig }    { \sqrt 2}    \int _{\R^3 } e^{  it H(g) }   \Big  ( a(k) \otimes [ E^{\star} (k) ,   X ]
 +  a^{\star}(k) \otimes [ E (k) ,   X ] \Big )
  e^{ - it H(g) } dk. $$

 \end{prop}

The proof of this Proposition is only a combination of the
Heisenberg equation for the
Pauli-Fierz Hamiltonian with the integral expression (\ref{Hint-Ek})  of $H_{int}$.

\begin{prop}\label{step-2}  For each $k$ in $\R^3$, we have:
$$ e^{i t H(g) }  (a(k) \otimes I)  e^{-i t H(g) } =
 e^{- i t | k| }  (a(k) \otimes I) - \frac {ig  }   {\sqrt 2 }
  \int _0^t e^{i (s-t)  | k|  }  e^{i s H(g) }  (I \otimes E(k) )
    e^{-i s H(g) } ds, $$
$$ e^{i t H(g) }  (a^{\star} (k) \otimes I)  e^{-i t H(g) } =
 e^{ i t | k| }  (a^{\star} (k) \otimes I) +  \frac {ig  }   {\sqrt 2 }
  \int _0^t e^{i (t-s)  | k|  }  e^{i s H(g) }  (I \otimes E^{\star} (k) )
    e^{-i s H(g) } ds. $$

\end{prop}

{\it Proof.} One knows that:
$$ e^{ i t H(0)    } ( a(k) \otimes I)  e^{ - i t H(0)    }=
  e^{ - i t |k|    } ( a(k) \otimes I).
  $$
Let:
$$ f(t)= e^{ i t |k|  }     e^{i t H(g) }  (a(k) \otimes I)  e^{-i t H(g) }
= e^{i t H(g) } e^{-i t H(0) } (a(k) \otimes I) e^{i t H(0) } e^{-i t H(g) }. $$
We see that:
$$ f'(t) = ig  e^{ i t |k|  }     e^{i t H(g) } [ H_{int } , (a(k) \otimes I) ]  e^{i t H(g) }. $$
According to the integral expression (\ref{Hint-Ek}) of $H_{int}$, since the operators $a(k')$ and $a(k)$ are commuting, and using $[ a(k), a^{\star } (k')]
= \delta (k - k')$, we get:
$$  [ H_{int } , (a(k) \otimes I) ]  = - \frac {1} {\sqrt 2}  ( I \otimes E(k) ). $$
Consequently:
$$  f'(t) = -  \frac {ig  }   {\sqrt 2 }  e^{ i t |k|  } e^{i t H(g) } ( I \otimes E(k) ) e^{-i t H(g) }.$$
The Proposition then follows. \fpr

The next Proposition is of the  type of Proposition  \ref{step-1}.

 \begin{prop}\label{step-3} For all operators $Y$ in ${\cal H}_{el } \otimes {\cal H}_{sp } $,
we have:
$$  e^{- i t H(g) } ( I \otimes Y) e^{ i t H(g) } =  I \otimes Y^{ free } (-t)
- ig \int _0^t e^{- i s H(g) }  [ H_{int} ,  ( I \otimes Y^{ free } (s-t) ) ]  e^{ i s H(g) } ds. $$

 \end{prop}

  {\it Proof.} For any operator  $Z$, set:
  $$ f(t) =  e^{- i t H(g) } e^{ i t H(0) }  ( I \otimes Z) e^{- i t H(0) }   e^{ i t H(g) }.  $$
 We have:
   $$ f'(t) =  - ig e^{- i t H(g) } [ H_{int} ,  ( I \otimes Z^{free}(t)  ) ] e^{ i t H(g) }.  $$
 Thus:
   $$  e^{- i t H(g) } ( I \otimes Z^{free}(t) )   e^{ i t H(g) } =
   (I \otimes Z) - ig \int _0^t
   e^{- i s H(g) }  [ H_{int} ,   ( I \otimes Z^{free}(s)  ) ] e^{ i s H(g) }  ds. $$
 The proof is completed  applying this equality to  $Z = Y^{free}(-t)$.\fpr

 {\it End of the proof of Theorem \ref{equation +reste}.} By Proposition
 \ref{step-1}, we have:
 $$ \frac {d}  {d t } e^{ it H(g) }  ( I \otimes X)
 e^{ - it H(g) } = i e^{ it H(g) } \Big ( I \otimes  [ (H_{el} +  H_{sp}), X ] \Big )
    e^{ - it H(g) } +    I_1(t) X + I_2(t) X, $$
    with:
 $$  I_1 (t) X = \frac { ig }    { \sqrt 2}   \int _{\R^3 } e^{  it H(g) }     ( a(k) \otimes [ E^{\star} (k) ,  X ])
  e^{ - it H(g) } dk, $$
 $$ I_2(t) X = \frac { ig }    { \sqrt 2}   \int _{\R^3 } e^{  it H(g) } \Big  (  a^{\star}(k) \otimes [ E (k) ,  X ] \Big )
  e^{ - it H(g) } dk. $$
  Proposition \ref{step-2} is used to rewrite $I_1 (t) X$.
This gives:
  $$  I_1 (t) X = \frac { ig }    { \sqrt 2}    \int _{\R^3 } e^{  it H(g) }  ( I \otimes [ E^{\star} (k), X ] )
  ( a(k) \otimes I) e^{ - it H(g) } = I'_1(t) X + I''_1(t) X, $$
  with:
  $$ I'_1(t) X = \frac { ig }    { \sqrt 2}  \int _{\R^3 } e^{  it H(g) }  ( I \otimes [ E^{\star} (k), X ] ) e^{ - it H(g) }
    e^{  it|k| }( a(k) \otimes I) dk, $$
   $$ I''_1(t) X =   \frac {g ^2 }   {2 }  \int _{\R^3 \times (0, t)} e^{  it H(g) }  ( I \otimes [ E^{\star} (k), X ] ) e^{i (s-t)  | k|   }  e^{i (s-t)  H(g) }  (I \otimes E(k) )
    e^{ - i s  H(g) } dk ds.  $$
Notice that $ I'_1(t) X  \sim 0$.  To rewrite $ I''_1(t) X$,
   we use Proposition \ref{step-3} with  $t$ remplaced by  $t-s$ and $Y$
   by $E(k)$. Thus, we obtain:
$$  e^{ i (s-t)  H(g) } ( I \otimes E(k) ) e^{ i (t-s)  H(g) } =  I \otimes E^{ free } (k, s-t) $$
$$   - ig \int _0^{t-s}  e^{- i s_2 H(g) } \Big [ H_{int} ,  ( I \otimes E^{ free } (k, s + s_2-t) ) \Big ]
 e^{ i s_2 H(g) } ds_2. $$
 Consequently:
   $$ I_1 (t) X \sim I''_1(t) X =  e^{  it H(g) } (I \otimes  L_1 (t)X)  e^{ - it H(g) }  + R_1 (t,g) X$$
 where $ R_1(t,g) X $ is defined in (\ref{R1-t}) and:
   $$ L_1(t) X =  \frac {g^2   }   { 2 }  \int _{\R^3 \times (0, t)}  e^{i (t-s)  | k|   }
    [ E^{\star} (k), X ]   E^{ f\ ree } (k, s-t)   dk ds. $$
   Next, we similarly consider $I_2(t)X$ and obtain:
    $$ I_2 (t) X \sim  e^{  it H(g) } (I \otimes  L_2 (t)X)  e^{ - it H(g) }  + R_2 (t,g) X,$$
   $$ L_2 (t) X = -   \frac {g^2 }   {2}  \int _{\R^3 \times (0, t) }
   e^{  i s |k |} E^ {free} ( k, -s) ^{ \star }  \  [ E(k), X]   dk ds $$
    where $R_2 (t,g) X$  is defined in (\ref{R2-t}).
   We see that the operator   $L(t, g) X$  defined in   (\ref{L-1-t}) satisfies:
    $$  L(t, g) X =  i [ (H_{el} +  H_{sp}), X ] +    L_1 (t) X + L_2 (t) X. $$
  The proof is completed. \fpr

 \subsection{ Norm estimates of the  Lindblad operator.}\label{s4.2}

We study here the main term
$L (t, g) X$
and its limit as $t\rightarrow +\infty$.
To this end, we need the next Proposition.

\begin{prop}\label{def-Z-t-X} Let $X$  be a bounded  operator in ${\cal H} _{el} \otimes {\cal H} _{sp}$
 which is also bounded  from  $W_1^{el}$  to itself.
Then for all $t>0$, the following operators are well defined from $W_2^{el} $ into $W_0^{el}$:
$$
    \Phi  (s , X) =  \int _{\R^3 }  e^{is|k| }  [ E (k)^{\star} , X]
    E ^{free}  (k, -s)   dk $$
$$
  \Phi ^{\star} (s , X) =   \int _{\R^3 } e^{-is|k| }  E^{ free} (k, -s)^{ \star}\  [X , E(k)] dk.  $$
Moreover, under the above hypotheses, there exists $C>0$ such that:
$$  \Vert   \Phi  (s, X)  \Vert
_{ {\cal L} ( W_2 ^{el} ,  W_0 ^{el} ) } + \Vert   \Phi ^{\star} (s, X)  \Vert
_{ {\cal L} ( W_2 ^{el} ,  W_0 ^{el}   ) }
 \leq  \frac { C}  { 1+ s^2}   \Vert X \Vert_{ {\cal L} ( W_1^{el}  , W_1 ^{el}   ) }.  $$
If in addition $X$ is bounded from $W_0^{el} $ to  $W_1^{el} $ then  we have:
$$  \Vert \Phi  (s, X)  \Vert
_{ {\cal L} ( W_1^{el} ,  W_0^{el} ) } + \Vert   \Phi ^{\star} (s, X)  \Vert
_{ {\cal L} ( W_1^{el} ,  W_0^{el} ) }
 \leq  \frac { C}  { 1+ s^2}   \Vert X \Vert_{ {\cal L} ( W_0^{el} , W_1^{el}   ) }.  $$
If the function $\varphi$ in   (\ref{A-j})(\ref{B-j})
is vanishing at the origin then the first  inequality can be replaced by:
$$  \Vert   \Phi  (s, X)  \Vert
_{ {\cal L} ( W_2 ^{el} ,  W_0 ^{el} ) } + \Vert   \Phi ^{\star} (s, X)  \Vert
_{ {\cal L} ( W_2 ^{el} ,  W_0 ^{el}   ) }
 \leq  \frac { C}  { 1+ s^3}   \Vert X \Vert_{ {\cal L} ( W_1^{el}  , W_1 ^{el}   ) }  $$
and similarly for the second inequality.

\end{prop}

{\it Proof. } We use spherical coordinates for $k$, $k= \rho \omega$,
$\rho >0$ and $\omega \in S^2$. We have:
$$
    \Phi  (s , X) =  \sum _{j=1}^3 \int _{\R_+ \times S^2   }  e^{is \rho  }
    [ F _j( \rho, \omega ), X]   G _j( \rho, \omega , s ) \rho  d \rho d \omega, $$
 with:
 $$  F _j( \rho, \omega ) = \rho ^{1/2}   E_j (\rho \omega )^{\star},\quad
   G _j( \rho, \omega , s ) =  \rho ^{1/2}   E_j ^{free}  (\rho \omega , -s).  $$
According to Proposition \ref{act-deriv-E-k}, we get:
 $$ \Vert \partial _{\rho}^{\alpha }   F _j( \rho, \omega ) \Vert
 _{ {\cal L} ( W_p^{el} , W_{p-1} ^{el} ) } \leq C ( 1+ \rho)^ {-2}.  $$
Using the hypotheses on the operator $X$, we obtain:
 $$ \Vert \partial _{\rho}^{\alpha }  [ F _j( \rho, \omega ) , X] \Vert
 _{ {\cal L} ( W_p^{el} , W_{p-1} ^{el} ) } \leq C ( 1+ \rho)^ {-2} \sup  _{m \leq p}
 \Vert X \Vert _{ {\cal L} ( W_m^{el} , W_m ^{el} ) }. $$
 By Proposition \ref{act-deriv-E-k} and since
  $ e^{ i t ( H_{el} + H_{sp}) }$ is uniformly bounded from
  $W_k^{el}$ into itself for each $k$, we get:
  $$ \Vert \partial _{\rho}^{\alpha } G_j ( \rho , \omega , s)  \Vert
 _{ {\cal L} ( W_2^{el} , W_1 ^{el} ) } \leq C ( 1+ \rho) ^ {-2}.$$
We can integrate twice by parts with the variable  $\rho$ in the expression of $\Phi (s, X)$. We deduce:
$$ s^2  \Phi (s, X)  = \sum _{j=1}^3 \int _{\R_+ \times S^2   }  e^{is \rho  }
     \partial _{\rho }  ^2
    \Big (  \rho  [ F_j (\rho, \omega) , X]  G_j ( \rho, \omega, s)  \Big )
    d \rho d \omega.  $$
To prove the last points of the Theorem, we note that if the function
 $\varphi$  vanishes at the origin then we can integrate by parts three times instead of two.

 The proof of the Proposition is completed.\fpr

We then deduce the next result.

    \begin{prop}\label{majo-R0}   Let $X$  be a bounded  operator in ${\cal H} _{el} \otimes {\cal H} _{sp}$
 which is also bounded  from  $W_1^{el}$  to itself. Then the operator
$L(t, g)X$ defined in (\ref{L-1-t}) is bounded uniformly in $t$
from $W_2^{el}$ to  $W_0 ^{el} $. Moreover, as  $t$
tends to  $+\infty$,  this operator  tends in
 $ { \cal L} ( W_2^{el}  , W_0 ^{el}  )$ to a limit $L(g) X$ satisfying:
$$ \Vert L (t, g) X - L(g) X \Vert
_{ { \cal L} ( W_2^{el}  , W_0 ^{el}  ) } \leq   \frac { Cg^2}  { 1+ t}
 \Vert X \Vert_{ {\cal L} ( W_1^{el} , W_1^{el}   ) }  $$
 and the above factor $1/(1+t)$ is improved to  $1/(1+t^2)$ if the ultraviolet cutoff $\varphi$ vanishes at the origin.
     \end{prop}

Propostion  \ref{defi-Lind} is then proved.

\subsection{ Estimates on the integral rests.}\label{4.3}

 We now need to control the norms in some spaces of the operators
 $ R_1(t,g) X$ and $ R_2(t,g) X$ given by (\ref{R1-t}) and (\ref{R2-t}).

  \begin{prop}\label{majo-R1}   For each integer $m$, there exists  $C_m>0$ such that
   the operator $R_1(t,g)$ defined in (\ref{R1-t}) satisfies:
  $$ \Vert R_1(t,g) X \Vert _{{\cal L} ( W_m^{tot} , W_{m-4}^{tot}) }
  \leq C_mg^3 \ln (1+t) \sup _{j\leq m} \Vert X \Vert
  _{{\cal L} ( W_j^{el} , W_{j}^{el }) }.$$
    We also have:
  $$ \Vert \sigma _0 R_1(t,g) X \Vert _{{\cal L} ( W_4^{el} , W_{0}^{el}) }
  \leq C_m g^3\ln (1+t) \sup _{j\leq 4} \Vert X \Vert
  _{{\cal L} ( W_j^{el} , W_{j}^{el }) }. $$
  If the smooth cutoff $\varphi$ in  (\ref{A-j}) and (\ref{B-j})
vanishes at the origin then the factor $\ln (1+t)$ can be omitted in the
above inequalities.

The same estimates holds true for $R_2(t,g)$ instead of $R_1(t,g)$.

     \end{prop}

 {\it Proof.} In the definition (\ref{R1-t}) of $ R_1(t,g)X $, we use spherical coordinates for  $k$, namely  $k = \rho \omega $ with $\rho >0$
and $\omega \in S^2$ the unit sphere centered at the origin.  Thus, we can write:

$$ R_1(t,g)X = g^3 \int _{  E(t) } \Phi (t, s_1, s_2) ds_1 ds_2 $$
with:
 \be\label{Phi-t}  \Phi (t, s_1, s_2) = \int _{ \R_+ \times S^2}  e^{ i (t - s_1) \rho  }
 e^{ it H(g) } \ ( I \otimes [ F( \rho , \omega ), X]  ) e^{ -i s_2 H(g)  }  $$
 $$   [ H_{int}  , I \otimes  G ( \rho , \omega , s_1, s_2, t)]
   e^{ i (s_2 - t)  H(g)  } \rho  d\rho d\omega  \ee
where:
$$ F( \rho , \omega ) = \rho ^{1/2}    E^{\star } (\rho \omega ) $$
$$ G ( \rho , \omega , s_1, s_2, t) =  \rho ^{1/2}   E ( \rho \omega , s_1 + s_2 - t ).  $$
According to Proposition \ref{act-deriv-E-k}, the function $F$ takes its values in ${\cal L } ( W_j^{el} , W_{j-1}^{el})$ and we have:
  $$  \Vert \partial _{\rho } ^{\alpha } F( \rho , \omega ) \Vert
   _{{\cal L} ( W_j^{el  } , W_{j-1}^{el }) } \leq C_j (1+ \rho) ^{-2}.
    $$
Under the hypothesis of  Theorem \ref{TP1}, we see:
  $$  \Vert \partial _{\rho } ^{\alpha }[ X ,  F( \rho , \omega )]  \Vert
   _{{\cal L} ( W_j^{el  } , W_{j-1}^{el }) } \leq C_j (1+ \rho) ^{-2}
   \sup _{k\leq j} \Vert X \Vert
  _{{\cal L} ( W_k^{el} , W_{k}^{el }) } .
    $$
   By Proposition \ref{norme-tenso}:
  $$  \Vert \partial _{\rho } ^{\alpha } ( I \otimes [ X ,  F( \rho , \omega )])  \Vert
   _{{\cal L} ( W_j^{tot  } , W_{j-1}^{tot }) } \leq C_j (1+ \rho) ^{-2}
   \sup _{k\leq j} \Vert X \Vert
  _{{\cal L} ( W_k^{el} , W_{k}^{el }) }.
    $$
In the same way and since  $ e^{ i t ( H_{el} + H_{sp}) }$ is uniformly bounded from
  $W_k^{el}$ into itself for all $k$, we deduce:
  $$ \Vert \partial _{\rho } ^{\alpha } G ( \rho , \omega , s_1, s_2, t)
  \Vert  _{{\cal L} ( W_j^{el } , W_{j-1}^{el }) }  \leq C_j (1+ \rho) ^{-2}  $$
  where $C_j$ is idependent on  $t, s_1 , s_2$. Therefore,  by Proposition \ref{norme-tenso}:
  $$ \Vert \partial _{\rho } ^{\alpha }  ( I   \otimes  G ( \rho , \omega , s_1, s_2, t)
  \Vert  _{{\cal L} ( W_j^{tot } , W_{j-1}^{tot  }) }  \leq C_j (1+ \rho) ^{-2}. $$
  By Proposition \ref{H-int-Sobol},  $H_{int } $ is bounded from  $W_{j}^{tot  }$ to $W_{j-2}^{tot  }$. Thus:
  $$  \Vert \partial _{\rho } ^{\alpha } [ H_{int}, I \otimes  G( \rho , \omega ,  s_1, s_2, t  ) \Vert
   _{{\cal L} ( W_j^{tot } , W_{j-3}^{tot }) } \leq C_j (1+ \rho) ^{-2}.  $$
Finally, we know that $ e^{ it H(g) }$, $e^{ -i s_2 H(g)  }$ and $e^{ i (s_2 - t)  H(g)  } $
   are uniformly bounded from $ W_j^{tot } $ into itself for all $j$.
   We can integrate twice by parts with the variable $\rho$ in    (\ref{Phi-t}). In view of the above estimates, we get:
    $$ ( 1 + |t - s_1|^2 ) \Vert  \Phi (t, s_1, s_2) \Vert
     _{{\cal L} ( W_j^{tot } , W_{j-4}^{tot }) } \leq C_j
     \sup _{k\leq j} \Vert X \Vert
  _{{\cal L} ( W_k^{el} , W_{k}^{el }) }. $$
  If the cutoff $\varphi $ vanishes at $0$, we have
 $ F( 0 , \omega ) = 0$ and $ G ( 0 , \omega , s_1, s_2, t) = 0$.
  In equality   (\ref{Phi-t}), we can integrate three times by parts and obtain:    
    $$ ( 1 + |t - s_1|^3 ) \Vert  \Phi (t, s_1, s_2) \Vert
     _{{\cal L} ( W_j^{tot } , W_{j-4}^{tot }) } \leq C_j
     \sup _{k\leq j} \Vert X \Vert
  _{{\cal L} ( W_k^{el} , W_{k}^{el }) }.$$
  If $E(t)$ is the set defined in (\ref{E-t}), we see that:
  $$ \int _{E(t)} \frac {ds_1 ds_2 } {1 + |t - s_1|^2  }  \leq C \ln (1+t),\quad  \int _{E(t)} \frac {ds_1 ds_2 } {1 + |t - s_1|^3  }  \leq C. $$
  The first point in the Theorem is then deduced.
The second one follows
  by Proposition \ref{norme-tenso}.
  \fpr

  {\it End of the proof of  Theorem  \ref{TP1}.} By Theorem
  \ref{equation +reste}, we have (\ref{equa-diff}) with:
  $$ R_0 (t,g)X = L(t, g)X - L(g) X,\quad  R(t,g) X = R_1 (t,g)X + R_2(t,g) X. $$
  The norm estimate of
 $ R_0 (t,g)X$ comes from Proposition \ref{majo-R0} and the norm estimate of $R_1(t,g)X$ and $R_2(t,g)X$ from Proposition \ref{majo-R1}.
\fpr

 \section{ Transition probabilities.}\label{s5}

We shall make explicit in this section the matrix of the operator $L(g)$
in a suitable basis, and in a particular case. In this Section, there is one  (spinless)
 electron ($Q=1$),  no fixed spin particle and no external magnetic field ($ A^{ext} = 0$). Thus:
$$ {\cal H}_{tot} = {\cal H}_{ph} \otimes {\cal H}_{el},\quad
{\cal H}_{el} = L^2 (\R^3)$$
Therefore  $H_{el}$  is the usual   Schr\"odinger operator:
$$ H_{el} = - \Delta + V. $$
We make the following assumptions.

{\it
 $(H'_1)$ The potential $V$ is a non negative real valued $C^{\infty }$ function  on $\R^3$,
 and there exists a real  $M>0$ and a  constant $C_{\alpha } >0$ for each multi-index $\alpha $,  such that:
$$ | \partial ^{\alpha } V(x) | \leq C_{\alpha } ( 1+ |x| ) ^{M- |\alpha |}. $$

$( H'_2)$ There exists  $\gamma >0$ such that:
$$ V(x) \geq \gamma  |x|^M. $$
}
The following result is proved in \cite{BH}.

\begin{prop} Under hypotheses  $(H'_1)$ and  $(H'_2)$ the operator $H_{el}$ initially defined on   ${\cal S} (\R^3)$ has a unique self-adjoint extension.   The spectrum of  $H_{el}$ is discrete.
It is the set of finite multiplicity eigenvalues $\mu_j$  going to $+\infty$. There exists an Hilbertian basis $(u_j)$ of ${\cal H}_{el}$ satisfying:
$$ H_{el} u_j = \mu_j    u_j. $$
Each $u_j$ belongs to the intersection of the spaces $W_m^{el}$. We have for all $m>0$:
$$ D( H_{el}^m ) = \{ u \in H^{2m} (\R^3),\  V(x)^m u \in L^{2} (\R^3) \}. $$

\end{prop}

Consequently, Hypothesis
$(H)$ of  Section \ref{s3} is satisfied if the above assumptions $(H'_1)(H'_2)$ hold.

 We can rearrange the eigenfunctions in such a way that the sequence  $(\mu_j)$ is non decreasing.

We next define the transition probabilities.
 Let  $\pi_{u_j}$  be the orthogonal projection in ${\cal H} _{el}$ on the vector space
spanned  by   $u_j$.

\begin{defi} For all unitary eigenfunctions  $u_j$ and  $u_m$ of $H_{el}$, and for each
 $t>0$, the  transition  probability of  $u_m $ to $u_j$ is defined by:
 \begin{align*}
 P_{ m j } (t ) &=\   \langle  (  \sigma _0 S (t ,g) \pi_{u_j}   ) u_m, u_m \rangle_{{\cal H} _{el}}  \\
 & = \  \langle( I \otimes \pi_{u_j}  ) e^{ - it H(g) } (\Psi_0 \otimes u_m) ,  e^{ - it H(g) }
 (\Psi_0 \otimes u_m)  \rangle _ {_{{\cal H} _{tot}}} \\
& =\   \langle   \sigma _0 S (t ,g) \pi_{u_j}   , \pi_{u_m}  \rangle _{HS }.
 \end{align*}
  where $\langle \, , \ \rangle_{HS} $ is the scalar product of two Hilbert-Schmidt operators
  in ${\cal H}  _{el}$.

  \end{defi}

These identities together with our main result leads us to make explicit the following matrix:
\be\label{M-m-j}   M _{mj } =  g^{-2}  \langle ( L(g) \pi_{u_j} ) , \pi_{u_m}  \rangle_{HS} =
   g^{-2}  \langle  (L(g) \pi_{u_j} )u_m ,  u_m  \rangle_{{\cal H} _{el}}. \ee

We now omit the subscript ${{\cal H} _{el}}$ in the scalar product of ${{\cal H} _{el}}$.

 We believe that this matrix is the generator of a semi-group of Markov matrices which
 perhaps will be a good approximation of the matrix $ P_{ m j } (t )$.

\begin{theo}\label{matrice}   The  $M_{mj}$  defined by (\ref{M-m-j}) satisfy:
\begin{align*}
 M_{mj} &= \pi \sum _{\alpha  = 1}^3 \int _{|k| = \mu_m  - \mu_j}
 | \langle  E_{\alpha } (k) u_m, u_j  \rangle |_{\R^3} ^2 \ d\sigma (k)
\quad  {\rm if}\quad \mu_j  < \mu _m  \\
M_{mj} &= 0 \quad  {\rm if}\quad    \mu_j  >  \mu_m    \\
M_{jj}&= -  \sum _{ k\not = j } M _{kj }\\
M_{mj }&= 0  \quad  {\rm if}\quad \mu_j = \mu_m\  {\rm and} \  j \not = m.
\end{align*}
\end{theo}

 Therefore $M_{mj} \geq 0$ if $\mu_j  < \mu _m$,
$M_{mj} \leq 0$  if $\mu_j  = \mu _m$ and $M_{mj} = 0$ if
 $\mu_j  > \mu _m$. We have, for each $j$:
 $$ \sum _{ m} M _{mj } = 0. $$
 We remark that these  properties are supposed to be verified by the infinitesimal
 generator of a Markov semi-group.  We do not know if the matrix  $M_{mj}$ is  indeed  the infinitesimal generator of a semi-group neither if we get a good approximation of the
 $ P_{ m j } (t )$. If this holds true then, up to this approximation, the transition probability of an initial electronic state to a higher electronic energy level state is zero.

{\it Proof.} By the  definition (\ref{L-1-t}) of $L(g)$, since $[ H_{el}   , \pi _{u_j} ]= 0$,
we have:
 $$ L(g) \pi _{u_j}  =
 $$
 $$    \frac {g^2 }   {2}
 \lim _{t\rightarrow + \infty }  \int _{\R^3 \times (0, t) }  \Big ( e^{ - i s |k |}
  [ E^ {\star } (k), \pi _{u_j}]\  E^{free} (k, - s)  -  e^{  i s |k |} E^ {free} ( k, -s) ^{ \star }  \  [ E(k), \pi _{u_j}]  \Big ) dk  ds.   $$
 We have, for all integers $j$ and $m$:
$$   \langle L(g) \pi _{u_j} , \pi _{u_m} \rangle_{HS} =  {\rm Tr}
\left ( \pi _{u_m} L(g) \pi _{u_j} \right ). $$
 By the  definition (\ref{E-k-t}), we have for the traces:
\begin{align*}
 {\rm Tr} \left(  [ E^ {\star } (k), \pi _{u_j}]\  E^{free} (k, - s)  \pi _{u_m} \right )
&= e^{is (\mu_m - \mu _j )} \left |  \langle E(k) u _m , u_j \rangle\right | ^2\\
& - e^{is \mu _j }\delta _{jm}  \langle e^{ -is H_{el}  } E(k) u_j  E(k) u_j  \rangle
\end{align*}
and
\begin{align*}
 {\rm Tr} \left (   E^ {free} ( k, -s) ^{ \star }  \  [ E(k), \pi _{u_j}] \pi _{u_m} \right ) &=
     e^{-is \mu _j }\delta _{jm}  \langle e^{ is H_{el}  } E(k) u_j  E(k) u_j  \rangle
\\
&  - e^{is (\mu_j - \mu _m )}   \left |  \langle E(k) u _m , u_j  \rangle \right | ^2.
\end{align*}
Therefore:
$$ g^2 A_{mj} =   \langle L(g) \pi _{u_j} , \pi _{u_m} \rangle _{HS} = g^2 ( A_{mj} + \delta _{jm} \lambda _j)$$
 with:
$$ A_{mj} =
 \lim _{t\rightarrow + \infty }  \int _{\R^3 \times (0, t) }
 \cos ( s ( |k| + \mu_j - \mu_m )) \    \left |  \langle E(k) u _m , u_j  \rangle \right | ^2   dk  ds.   $$
We know that, for each suitable  function $F$,  we have, if  $\lambda < 0$:
$$ \lim _{t \rightarrow + \infty }
\int _{\R^3\times (0, t)  }  \cos ( s(|k| + \lambda ) ) \
F (k) dk dt = \pi \int _{ |k| = - \lambda } F(k) d \sigma (k).$$

This limit is zero if $\lambda \geq 0$. We then indeed obtain the expression of $M_{mj}$ for $m\not= j$. For the diagonal elements  $M_{jj}$, the above computations are not used but rather  $L(g) I= 0$, denoting by  $I$ the identity operator  on  ${\cal H}_{el}$. Since $I$ is the sum  of the  $\pi_{u_j}$,  it follows that
 $\sum _j \langle L(g) \pi_{u_j} , \pi_{u_m} \rangle _{HS} = 0 $  for each  $m$.
This completes the proof of  Theorem \ref{matrice}. \fpr

    laurent.amour@univ-reims.fr\newline
Laboratoire de Math\'ematiques de Reims UMR CNRS 9008,\\ Universit\'e de Reims Champagne-Ardenne
 Moulin de la Housse, BP 1039,
 51687 REIMS Cedex 2, France.

jean.nourrigat@univ-reims.fr\newline
Laboratoire de Math\'ematiques de Reims UMR CNRS 9008,\\ Universit\'e de Reims Champagne-Ardenne
 Moulin de la Housse, BP 1039,
 51687 REIMS Cedex 2, France.

\end{document}